\documentclass[doublecol]{epl2}
\usepackage{graphicx}
\usepackage{amsmath,amssymb}
\def\be{\begin{equation}}
\def\ee{\end{equation}}
\def\S{{\mathbb S}}

\title{Restricted equilibrium ensembles: Exact equation of state of  a model  glass }

\author{Deepak Dhar\inst{1} \and Joel L. Lebowitz\inst{2}}
\shortauthor{Dhar and Lebowitz}
\institute{
\inst{1}Department of Theoretical Physics, \\
Tata Institute of Fundamental Research, \\
Homi Bhabha Road, Mumbai-400005, India\\

\inst{2}Departments of Mathematics and Physics, \\
Rutgers University, Piscataway, \\
NJ 08854, USA.
}
 
\date{\today}

\pacs{05.70.Ce}{Thermodynamic functions and equations of state}
\pacs{64.60.My}{Metastable phases }
\pacs{02.50.Ga}{Markov processes}

\abstract{ 
We investigate the thermodynamic properties of a toy model of glasses: a 
hard-core lattice gas with  nearest neighbor interaction in one dimension. The time-evolution is Markovian, with nearest-neighbor and next-nearest neighbor hoppings, and the  transition rates are assumed to satisfy detailed balance condition, but  the system is
non-ergodic below a glass temperature.  Below this temperature, the system is in restricted thermal equilibrium, where both the number of  sectors, and the number of accessible states within a sector grow exponentially with the size of the system. Using partition functions that sum only over dynamically accessible states within a sector, and then taking a quenched average over the sectors,  we determine the exact equation of state of this system.  
}

\begin{document}

\maketitle

The glassy state is usually considered a non-equilibrium state of matter 
in the sense that the conventional Boltzmann-Gibbs treatment in terms of 
partition functions would not work, and yields very different properties 
of the corresponding ``equilibrium state" (e.g. of crystalline quartz, 
and not of window glass).  In this paper, our aim is to develop an equilibrium statistical 
mechanical description of the glassy state using restricted phase-space ensembles.  
We consider a system constrained by dynamics to a restricted region of phase space. 
Within this region, it acts like an equilibrium system. This may be considered as 
an idealized description for metastable states such as supercooled liquids, or a glass.
Focussing on the latter, we note that   materials like window glass, for a 
given history of preparation, and at temperatures sufficiently below the 
glass temperature, have a well-defined macroscopic density,  velocity 
of sound, and specific heat. Then, over a time scale of microseconds 
to years, these materials are in some effective restricted 
thermodynamic equilibrium.

 In the restricted equilibrium ensemble corresponding to a glassy state, the 
partition function sum only extends over a restricted region of the phase space. 
As an illustrative example of  these  ensembles, we discuss a simple  model. 
In particular, we determine 
the exact equation of state (a material -dependent relation between the 
density, pressure and temperature). To the best of our knowledge, this 
is the first time the exact equation of state for  a model  with 
short -range interactions and  showing a glassy phase, is obtained.
 
  The model we study  is a  lattice gas in one dimension, with a pair 
potential. The time evolution is assumed Markovian, with detailed 
balance, but in the glassy regime, the dynamics is non-ergodic, and 
phase space breaks up into a large number of disconnected sectors. We 
calculate averages of physical quantities in terms of partition 
functions within a sector, and then average this over sectors.

The notion of restricted ensembles is not  new.  Many authors have discussed metastable states in terms of constraints on the regions of phase space available to the system, i.e. phase space is broken into disjoint components, with ergodicity within components\cite{penrose}. The idea of components, in 
the context of supercooled liquids and glasses was made more specific as 
inherent structures by Stillinger and Weber \cite{inherent_structures}. 
This gives the energy or free-energy landscape picture, which has been 
very useful in providing a conceptual framework for the study of glasses. 
But the actual calculation of the partition function within a component, or the number of components is very difficult, and one usually just postulates the form of the 
distribution function of these \cite{multiplicity}. The study of 
specific models with Markovian evolution with detailed balance, showing 
break-down of phase space into disjoint sectors, was pioneered by 
Fredrickson and Andersen\cite{fredrickson}, and such kinetically 
constrained models have been studied a lot recently \cite{ritort}. 
However, usually the cases studied have a trivial Hamiltonian, and the 
main focus is on the relaxational dynamics. Our focus here
is on the  thermodynamics in the glassy phase.

We should also mention earlier work on metastable states  for systems with long-range 
couplings, as in the Sherrington-Kirkpatrick model of spin-glasses 
\cite{parisi}, or supercooled liquids \cite{glass, cavagna}. Theories 
like the mode-coupling theory of glasses mainly discuss the onset of 
glassiness from the liquid side \cite{mode-coupling}. An overview of  recent work 
on supercooled liquids and structural glasses may be found  in \cite{ediger, pnas}. 

Our approach uses pico-canonical ensembles, discussed earlier in \cite{irreducible}. The glass transition is built into the system by assuming temperature-dependent rates of local dynamic processes, some of which are set to be exactly zero in the glassy phase. We do not try to describe the  approach to the glass transition, but focus on the near-equilibrium behavior away from  the glass transition point, in the glassy phase.

\section{Definition of the model}

The system we study is a hard-core lattice gas on a semi-infinite line 
of sites, labelled by positive integers $i$, with $i$ varying from $1$ 
to $\infty$. At each site $i$, there is an occupation number variable 
$n_i$ that takes values $0$ or $1$, depending on whether the site is 
occupied or not.

There are a total of $N$ particles on the line. On the left, there is an 
immovable wall at $i=0$.  To the right of the  rightmost ( the $N$-th) particle, is a movable piston, whose position will be denoted by $L$, 
and which exerts a constant pressure $p$ on the system.  There is an 
attractive interaction between nearest neighbor occupied sites of 
strength $J$.  For simplicity, we assume that the 
interaction between piston and particles of the gas, and between the 
left wall and the particles is the same as that between two particles. Then,
with the convention $ n_0 = n_L =1$, the Hamiltonian of the system is given by 
\be H = - J 
\sum_{i=0}^{L-1} n_i n_{i+1} + p L \label{eqh} 
\ee 
We assume that the system evolves by Markovian dynamics, with the 
following rules:\\
 i) An occupied site with an empty neighbor can exchange position with 
the neighbor with a rate $\Gamma_1 \exp( -\beta \Delta E/2)$.  Here 
$\Delta E$ is the difference of energies between the final and initial 
configurations, and $\beta$ is the inverse temperature. This is 
represented by the `chemical' equation $ 01 \rightleftharpoons 10 $.
 
 ii) An occupied site with two empty neighbors on its left ( or right), 
can jump two spaces to the left ( right), with a rate $\Gamma_2 \exp(- 
\beta \Delta E/2)$. This is represented by the equation $ 100 
\rightleftharpoons 001 $.

 iii) The piston can exchange position with a neighboring empty site. 
This process is represented by the equation $ 0P 
\rightleftharpoons P0$. The corresponding rate is $\Gamma_3 exp( -\beta \Delta E/2)$.

In our model, the externally controlled variables are the pressure, and 
temperature. The dynamics conserves the number of particles.   Within the model, we are free to prescribe any  functional  dependence to the rates 
$\Gamma_1$, $\Gamma_2$ and $\Gamma_3$, and the long-time steady   state of the system, and the thermodynamic properties  do not depend on the precise values of these parameters.
We postulate that  $\Gamma_2$ and $\Gamma_3$ are 
non-zero for all temperatures, and may be assumed to be independent of 
temperature and pressure, without any loss of generality.  For $\Gamma_1(p,T)$,  we assume that in two-dimensional control space $(p,T)$, there are  
regions where it is zero, and in the rest of space, it is 
not [ fig. 1]. The regions where $\Gamma_1(p,T)$ is zero will be called glassy regions.

\begin{figure}

\begin{center} \includegraphics[width=6.0cm]{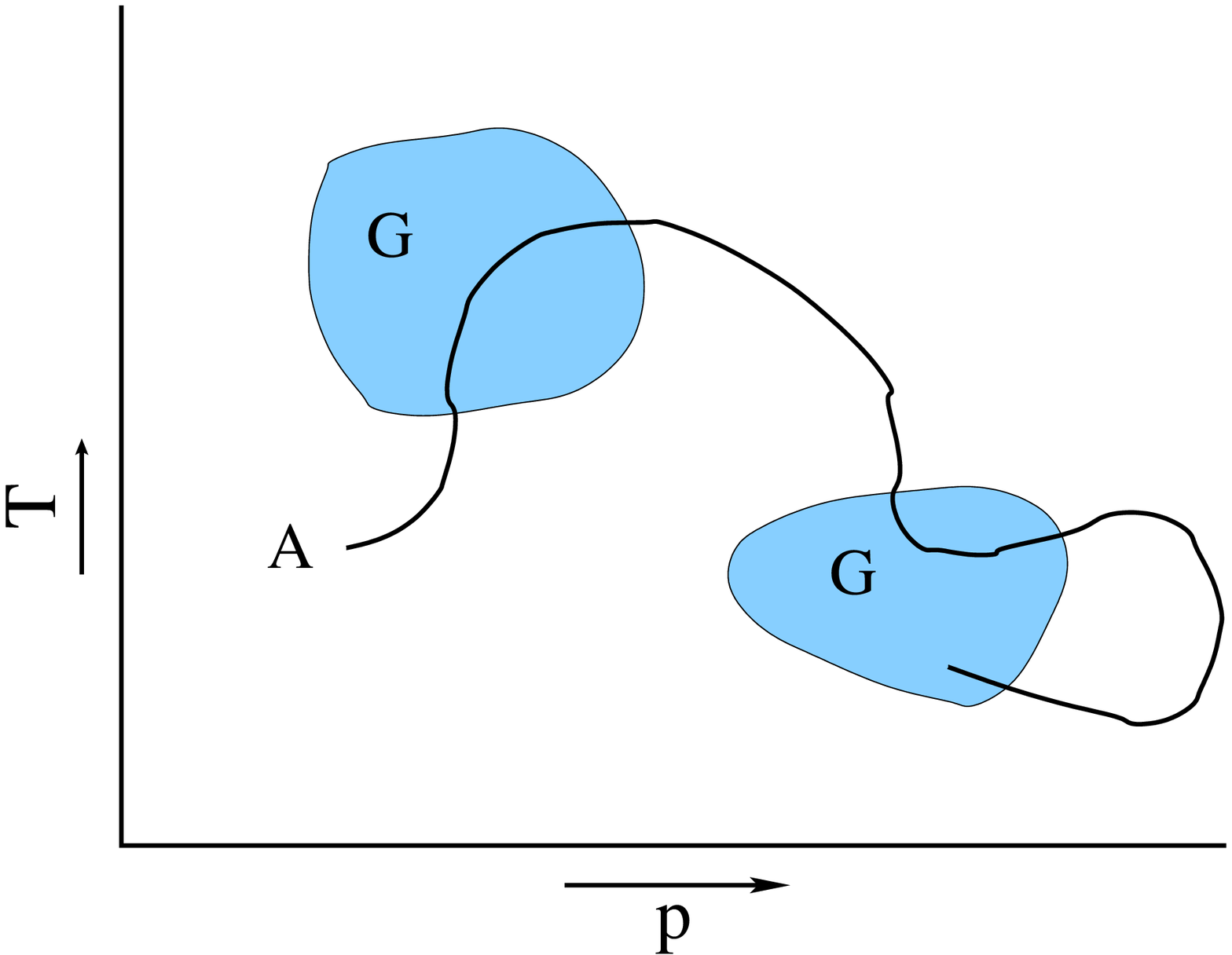} \caption{A schematic representation of  the  glassy and non-glassy regions of the control parameter space $(p,T)$. The 
non-ergodic glassy regions are shown as shaded regions $G$.  $A$ denotes a path in this space, corresponding to particular preparation of the system. } 
\end{center}
\label{fig1}

\end{figure}

The dynamics of this model is somewhat similar  to the model of diffusing reconstituting dimers (DRD model) discussed earlier in \cite{irreducible}.  In 
the DRD model, the only allowed transitions in the low-temperature phase are 
$110 \rightleftharpoons 011$, while in the high temperature phase, $01 \leftrightharpoons 10$ is also allowed.  These moves are the same as the moves in 
the present case if the $0$'s and $1$'s are interchanged. 

The DRD model was defined in the constant $L$ ensemble, and the right wall 
could not move. It is not possible to define a satisfactory dynamics of a piston 
consistent with the dynamics of the DRD model. For example, consider the 
configuration $1000110011000100P$. Then, it is easily seen that in the DRD dynamics, the piston cannot move to the left of the present position of the rightmost $1$. 
Hence the DRD lattice gas is incompressible in the non-ergodic phase.

Again, if we move the piston in the DRD model to the right by some 
macroscopic distance,  only dimers can move into 
the new region.  Thus, the long-time steady state is inhomogeneous, and shows a phase separation, with the left region at a higher density than the right, and  pressure equilibration between different parts  is not possible under the DRD dynamics. This makes  the DRD model  unsuitable for studying the equation of state. 
The present model does not suffer from  these  problems.

\section{Sector decomposition and the restricted ensemble}

We note that each of the Markov rates here satisfies the detailed 
balance condition, and  the  state with the probability of a 
configuration $C$ is given by the standard Boltzmann -Gibbs weight 
\be 
{\rm Prob }(C) = \exp[ - \beta H(C)]/{\cal Z}( p,T) 
\label{eq2}
\ee 
is a steady state. Here ${\cal Z}(p,T)$ is the partition function of the system, given by 
\be 
{\cal Z}(p,T) = \sum_{C} \exp[- \beta H(C)] 
\ee
If $\Gamma_1(p,T)$ is non-zero, any configuration of $N$
particles can be reached from any other. Then,  the measure given by Eq.  (\ref{eq2}) is the unique long-time steady state measure.

If however, $\Gamma_1(p,T) =0$, then the phase space breaks up into a large 
number of disconnected parts, called components, or sectors. There are $2^N$ disjoint 
sectors. This is easily seen: Define $x_i $ as the position of the 
$i$-th particle., and $\Delta_i = x_{i}- x_{i-1}$. Then the different 
$\Delta_i$ can only change by $\pm 2$. We define $\eta_i = \Delta_i (mod 
2)$. Then $\{\eta_i\}, i= 1$ to $N $ are conserved. Note that 
$\eta_{N+1}$ is not conserved. Clearly, the number of sectors is $2^N$. 
We will denote a sector by $\S$, and label these sectors by specifying 
the $N$ binary integers $\{\eta_i\}$.  

The conditional probability of a 
configuration $C$ in the long-time steady state, given that the system 
is in the sector $\S$ is given by
\be 
{\rm Prob}(C|\S) = \exp[ - \beta H(C)]/ {\cal Z}_{\S}(p,T). 
\ee 
The normalization factor ${\cal Z}_{\S}(p,T)$, called the pico-canonical 
partition function for the sector $\S$, is defined as the sum of the 
standard Boltzmann weights, but only over configurations $C$ that are in 
the sector $\S$, 
\be 
{\cal Z}_{\S}(p,T) = \sum_{ C \in \S} \exp[ - \beta H(C)] 
\ee

  We consider the control parameters $(p,T)$ being varied slowly along 
some curve $A$ in the control parameter space (Fig. 1). 
The curve $A$ may enter and leave the glassy regions several times. When 
the system is in a non-glassy region, there is a unique equilibrium 
state of the system, that is independent of the system history, and only 
depends on the current value of the control parameters. In the glassy 
regions, the restricted equilibrium state depends on $(p,T)$, and also 
on the sector. If we vary $(p,T)$, staying within the glassy region, the 
sector does not change. The relative probabilities of different sectors depend 
only on the point from which one entered the glassy region.

A precise specification of the sector $\S$ requires $N$ bits of 
information, and if $N$ is large, this is neither possible, nor useful. 
In an experimental set up, $\S$ is specified in some general way in terms of 
how the system is prepared.  We assume that in the beginning, the system is 
prepared in the ergodic region of the control-parameter space $(p,T)$, and then 
brought to the desired state by moving along a specified path $A$ adequately slowly.
We will specify the macroscopic state of system 
in a glassy state by the last values of $( p, T)$ in its history, when it 
was ergodic with $\Gamma_1 \neq 0$, to be denoted by $( p^*, T^*)$.  Then, 
the probability that the system falls in the sector $\S$ is given by 
\be 
{\rm Prob}(\S) = {\cal Z}_{\S}( p^*,T^*)/{\cal Z}(p^*,T^*) 
\ee

For any observable ${\cal O}$, the long-time average value within the 
sector $\S$ will be denoted by $\langle {\cal O} \rangle_{\S}$. 
This is given by
\be
\langle {\cal O}\rangle_{\S} =  \sum_{ C \in \S} {\cal O}(C) {\rm Prob}(C|\S)
\ee

Often, these can be expressed as appropriate derivatives of the free-energy, as in the standard 
equilibrium statistical mechanics. For example the average 
length $\langle L \rangle_{\S}$ of the system in the sector $\S$ is 
given by 
\be 
\langle L \rangle_{\S} = -k_B T \frac{\partial}{\partial p} 
\log {\cal Z}_{\S}( p,T). 
\ee
The expected value of an observable $\langle {\cal O} 
\rangle$ in an experiment, for a given history of preparation of  the sample, is obtained 
by further averaging $\langle {\cal O} 
\rangle_{\S}$ over different sectors $\S$ 
\be 
\langle {\cal O} \rangle = \sum_{\S} {\rm Prob}(\S) \langle {\cal O} \rangle_{\S}, 
\ee
where ${\rm Prob}(\S)$ is the probability that the system will freeze into the sector $\S$, for the given history of preparation.  

One can similarly calculate variance of the observable, or the sector-to-sector variation of the sector-mean $\langle {\cal O}\rangle_{\S}$.   
In the simple case we are considering, macroscopic quantities like the mean density can be shown to be self-averaging, and the variance of the sector-means are of ${\cal O}(N)$. 
The fluctuations, compared to the mean, are smaller by  a factor  $N^{-1/2}$, for large $N$. If the relative fluctuations are not small for some observable, predicting its value in  a particular experimental set-up, without  any additional information about the experiment, is  clearly not possible. In such cases,   one can only determine the probability distribution of such an observable.

In particular, the mean free energy in the glass phase is defined as
the  free energy, averaged over  sectors. We are working in the constant $(p,T)$-ensemble, and the appropriate free energy is the Gibbs free energy $\Phi(p,T)$. The average value of $\Phi(p,T)$ in the glass phase, averaged over sectors, will be denoted by $\langle \Phi_G(p,T) \rangle$.  This is defined as
\be
\langle \Phi_G (p,T) \rangle=  - k_B T  \sum_{\S} {\rm Prob}(\S) \log {\cal Z}_{\S}(p,T)
\ee
Note that  $\langle \Phi_G (p,T) \rangle$  implicitly depends also on $(p^*,T^*)$. 

\section{Equation of state}

 We now determine the equation of state for this system.  Consider first 
the ergodic case when $\Gamma_1 \neq 0$.  In this case, any allowed 
configuration can be reached from any other under the Markovian dynamics 
of the system.

We define $x_{N+1}$ to be the position of the piston $L$. In terms of 
the variables $\{\Delta_i\}$, the Hamiltonian can be written as

\begin{equation} H = \sum_{i=1}^{N+1} [-J \delta_{ \Delta_i,1} + p 
\Delta_i]. 
\end{equation} 
Clearly, in this case, the different $\Delta_i$ 
are independent random variables, we get the partition function of the 
system ${\cal Z}(\beta, p)$ in the constant temperature and pressure 
ensemble as \cite{takahashi}

\begin{equation} {\cal Z}(p, T) = w(u,x)^{N+1}, \end{equation}
where we have used the notation $u = e^{\beta J}$, and $x = e^{-\beta p 
}$, and
\begin{equation} w(u,x) = \sum_{\Delta = 1}^{\infty} u^{ 
\delta_{\Delta,1}} x^{ \Delta}= u x + \frac{x^2}{1-x}. \end{equation}
The mean spacing between the molecules $\langle \Delta \rangle$, which 
is the inverse of the density of the lattice gas, is given by

\begin{eqnarray} \nonumber \langle \Delta \rangle &=& x 
\frac{\partial}{\partial x} \log w(u,x)\\ &=& \frac{x}{w(u,x)} \left[ u 
+ \frac {2 x - x^2}{(1 - x)^2}\right], {\rm ~ for ~} \Gamma_1 \neq 0,~ 
\label{eqstate1} \end{eqnarray}
and the mean energy per particle $\langle E(u,x) \rangle$ is given by 
\begin{equation} \langle E(u,x) \rangle = -J u  \left[ u  + 
\frac{x}{1-x}\right]^{-1}, {\rm ~ for ~} \Gamma_1 \neq 0. 
\end{equation}

We can similarly calculate ${\cal Z}_{\S}(p,T)$, for any given sector 
$\S$. Define \begin{equation} w_{odd} = \sum_{\Delta ~odd} 
u^{\delta_{\Delta,1}} x^{\Delta} \end{equation}
and a similar equation for $ w_{even}$, where the sum is only over even 
values of $\Delta$. Then we have \be w_{odd} = ux + \frac{x^3}{1 - x^2} 
\ee and \begin{equation} w_{even} = x^2/( 1- x^2) \end {equation}

We denote the number of zero $\eta_i$'s by $r(\S)$.  It is easily seen 
that \be {\cal Z}_{\S} = w_{even}^{r(\S)} w_{odd}^{N -r(\S)} ( w_{odd} 
+ w_{even}) \ee

Under quenching, $\eta_i$ takes the value $0$ with probabilities 
$w_{even}/w = \alpha^*$ independent of $i$, and the value $1$ with 
probability $1 -\alpha^*$. 
\be 
\alpha^* = \frac{x^*}{x^* + {x^*}^2 + u^* ( 1 - {x^*}^2)} 
\ee 
Then, the probability of a sector $\S$ with exactly $j$ 
odd $\eta_i$'s, is given by \be {\rm Prob }(\S) = {\alpha^*}^{j} {( 1 - 
\alpha^*)}^{ N -j} \label{eqsfr} \ee
The variable $r(\S)$ is distributed as a binomial distribution \be {\rm 
Prob}(r(\S) = j) = ^{N}C_{j} ~{\alpha^* }^j ( 1 - \alpha^*)^{ N - j} 
\label{eq22}\ee 
Thus the distribution of $r(\S)$ is sharply peaked, with maximum at $
\alpha^* N$, with a width of order ${\cal O}( N^{1/2})$.

A straight forward calculation gives the mean spacing between the 
particles in the low-temperature phase as 
\be 
\langle \Delta \rangle = x 
\frac{ \partial}{\partial x}[ \alpha^* \log w_{even}(u,x) + ( 1- 
\alpha^*) \log w_{odd}(u,x) ]. 
\ee 
Using the expressions for $w_{odd}$ 
and $w_{even}$ (Eqs. 12 and 13), we finally get 
\begin{eqnarray} 
\langle \Delta \rangle=( 1 - 
\alpha^*) &\left[ 1 + \frac{ 2 x^2 ( 1 - x^2)^{-2}}{  u + x^2 ( 1 - x^2)^{-1}}\right]+\alpha^* \frac{2}{(1 - x^2)}, \nonumber \\
  & {\rm ~~~~~~~~~~~~~~~~~~~~~~~~~for ~} \Gamma_1 = 0.
\label{eq23}
\end{eqnarray}

This equation, along with Eq. (\ref{eqstate1})gives the mean volume per 
particle as a function of $u$ and $x$, and hence is the equation of 
state of the material.  It depends on the history of the system through 
the parameter $\alpha^*$, which depends on onset of the glassy state. From Eq.(\ref{eq22}), the sector-to-sector variation of $r(\S)/N$ is of ${\cal O}(N^{-1/2})$, and hence fluctuations in $\langle \Delta \rangle_{\S}$ are also small for large systems. 

Similarly, we can calculate the mean energy per particle $\langle E(u,x)\rangle $ by taking derivatives with respect to $u$, and we get 
\be \langle E(u,x) \rangle  = \frac{ - J u ( 1 - \alpha^*)}{ 
u + x^2/(1 - x^2)}, {\rm ~ for ~} \Gamma_1 = 0.~~ \ee

As an example, consider the simple case $ J=0$. In this case, there is no energy, and 
the density is a function of only one variable $\beta p$. In this case, it is easily seen that the 
equilibrium density is given by \be \rho_{eq} = 1 - e^{ -\beta p} \ee.

We assume that the high-density phase is not ergodic,  say whenever $\rho > 1 -  x^*$. 
Then, it is easy to see from Eq.(\ref{eq23}), that  
in the non-ergodic phase, the dependence is 
\be \rho_{glass} = ( 1 
- e^{-2 \beta p}) \left[ 2 \alpha^* + ( 1 - \alpha^*)( 1 + e^{-2 
\beta p})\right]^{-1} \ee where $\alpha^* = x^*/( 1 + x^*)$. 

\begin{figure} \includegraphics[width=8.0cm, angle =0]{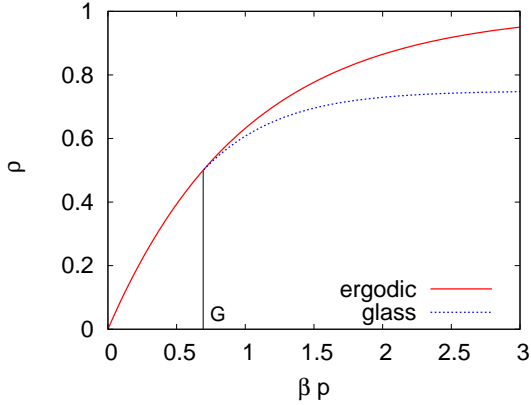} 
\caption{Plot of density $\rho$ as a function of the scaled pressure $\beta p$ for $J=0 $, when  the regime $\beta p^* \geq \log(2)$  is assumed to be non-ergodic. The point of onset of non-erdodicity is denoted by $G$ in the figure.}
\label{fig:2}

\end{figure}

We have plotted $\rho_{eq}$ and $\rho_{glass}$ in Fig. \ref{fig:2} for $x^* = 1/2$. We note that the density is continuous at
$ p = p^*$, and the glass is less compressible than the corresponding equilibrium state. 

\section{Entropy}

The statistical mechanical entropy of a macroscopic system in equilibrium, i.e.
one described by the Gibbs ensemble,  when the state state is discrete, and the Gibbs ensemble assigns  probability $p_i$ to the $i$th state, is given by
\be
S = -\sum_{i} p_i \log p_i,
\label{eq28}
\ee
This entropy coincides, to leading order in the size of the system, with the thermodynamic entropy of Clausius,  and is the same for different ensembles, for ordinary systems ( we are ignoring  gravitational interactions).  The set of microstates used is always a restricted one, in practice. For example, when considering argon gas below $10^{3}~ ^{\circ}$K, we do not consider the ionized states, or the excited nuclear states.  There is some controversy in the glass literature \cite{controversy} about the appropriate choice of the set of microstates over which one should sum in  Eq.(\ref{eq28}) . In the following, we shall assume that the sum extends over all states
which could be reached, given the thermal history of the system. There are $^{L}C_N$ for each $L$, and $L = N+1, N+2, \ldots.$

 This entropy can be expressed as a 
sum of two terms.  Firstly, we do not know which sector the system has 
fallen into. The corresponding entropy, usually called the {\it frozen 
entropy} of the system, and is given by \be S_{frozen} = - \sum_{\S} 
{\rm Prob}(\S) \log {\rm Prob}(\S) \ee
This is easily computed in our case using Eq.(\ref{eqsfr}), and we get 
frozen entropy {\it per particle} $ s_{frozen}$, in the thermodynamic 
limit of large $N$ given by
\be s_{frozen} = - \alpha^* \log \alpha^* - (1 - \alpha^*) \log ( 1 - 
\alpha^*) \ee
 For $ x^* = 1/2$, we have  $\alpha^* =1/3$, and the frozen entropy per particle is $ \log 3 - \frac{2}{3} \log 2 
\approx 0.636$, which is  a large fraction of the total entropy per particle at the glass point $s^* =
 \log(4) \approx 1.386$. 
 
Note that $s_{frozen}$ only depends on $ ( u^*, x^*)$, and not on $u$ or 
$x$ explicitly. Hence in taking derivatives with respect to $u$ or $x$, 
it does not contribute, and the mean energy or pressure are same as 
would be computed from pico-canonical partition function.

The second contribution to entropy comes from the many possible 
microstates within one sector. This depends on the sector, and its 
calculation involves the pico-canonical ensemble. This will be denoted 
by $S_{pico}$, and is given by 
\be S_{pico}(\S) = - \sum_{C \in \S} {\rm Prob}( C | \S) 
\log {\rm Prob}( C | \S) \label{eqspico} 
\ee
This can be expressed in terms of the picocanonical partition function, 
for large $N$, to leading order in $N$, as \be S_{pico}(\S) = \log {\cal 
Z}_{\S} - \beta \frac{\partial}{\partial \beta} \log {\cal Z}_{\S} \ee

Let $f_j$ is the conditional probability that a particular $\Delta$ 
takes the value $j$, given whether $j$ is even or odd. Clearly, $f_1 = u 
x/ w_{odd}$, $f_{2 j + 1} =x^{2 j + 1}/ w_{odd}$ for $j >1$, and $ f_{2 
j} = x^{ 2 j-2} ( 1 - x^{2})$. Using the fact that ${\rm Prob}(C | \S)$ 
is a product of probabilities of different $\Delta_i$'s, it is 
straightforward to calculate the {\it per site} entropy, $s_{pico}$, 
averaged over the sectors $\S$, and we get \be s_{pico} = -\alpha^* 
\left[\sum_{j~ even} f_j \log f_j\right] -( 1 - \alpha^*) \left[\sum_{j 
~odd} f_j \log f_j \right] \ee

The controversy in literature about the entropy of glasses, in our model,
relates to the question whether to include $s_{frozen}$ in the thermodynamic entropy. 
This seems to us  a matter of taste. One can argue that if we start from the high
temperature phase, and cool down along some annealing path $A$ (Fig. 1), then the microstates in different sectors should contribute to entropy. On the other hand, if
we prepare the system in some sector $\S$, then other states not consistent with this
initial preparation would not contribute. It is clear is that $s_{frozen}$ will not change  if we change control parameters staying  within the glassy phase.  This could
be a change of temperature in the glassy phase, and  whether we add $s_{frozen}$ or not, we would get the same specific  heat in the glassy phase.  

We note finally that  the 1-d lattice-gas model with nearest neighbor couplings 
can be seen as a renewal process, i.e. the probability that separation between the $i$th particle and $(i+1)$th particle being $j$ is independent of $i$. Denote this by $c_j$.  A straightforward calculation gives 
\begin{eqnarray} c_j 
&=& f_j \alpha^* , {\rm ~for ~} j ~even;\\ & =& f_j ( 1 -\alpha^*) , {\rm 
~for~} j ~odd. 
\end{eqnarray}
With this distribution, the total entropy per site is given by \be s 
=s_{frozen} + s_{pico} = - \sum_j c_j \log c_j. \ee

\section{Discussion}

One important drawback of this particular model deserve mention. It does 
not have a crystalline phase at low temperatures, and consequently there 
is also no supercooled liquid phase. In the model, there is a direct 
transition from the ergodic liquid to the non-ergodic glass phase.

Of course, $\Gamma_1$ becoming exactly zero in the glassy regime is an 
idealization, used to make the problem more tractable. It is not 
qualitatively different from other idealizations, such as infinite heat 
reservoirs with weak couplings etc. that are routinely used in 
equilibrium statistical mechanics.

One can ask how does the relaxation rate between different sectors 
depend on $\Gamma_1$, when $\Gamma_1$ is very small.  In this case, if 
one looks at the process as a change in the sector-labels $\{\eta_i\}$, 
with time, the process is like a symmetric exclusion process in this 
space, where $01 \rightleftharpoons 10$ with rate $\Gamma_1$. From 
general arguments, the relaxation time of this process should grow as 
$L^2/\Gamma_1$.

  In general, for a non-equilibrium system, with a Hamiltonian
$H(\lambda)$, in contact with a heat bath at a fixed temperature $T$, if
a control parameter in the Hamiltonian is changed from $\lambda$, by a
small amount to $\lambda+ \Delta \lambda$, one can write $\Delta \bar{E}$, 
the change in the
mean energy $\bar{E}$ of the system as 
\begin{equation} \Delta \bar{E} =
\Delta \sum_i p_i E_i =\sum_i \left[ p_i \Delta E_i + E_i \Delta p_i
\right].
\end{equation}
Here the first term may be identified as the work
done on the system $\Delta W$, and the second as the amount of heat
added $\Delta Q$ \cite{jarzynski}. In our model, if we go across the
glass transition, neither $p_i$, nor $E_i$ undergo any change. Hence,
there is no work done on the system, or heat exchanged with reservoir,
as we cross the ergodic-glass boundary.

It is quite straightforward to generalize this model. For example,  
instead of step-size $2$, we can consider step size $3$, or $4$ etc..
We can also consider longer-ranged interactions. This does not change the sector 
decomposition, but calculation of partition function within a specified 
sector becomes  more complicated, but still reducible to a finite matrix diagonalization, if the range of interaction is finite.

We can also construct models in higher dimensions. 
Consider, for example, a lattice gas on a square lattice, with a 
pair-wise additive interaction potential of finite range. At high 
temperatures, we assume that each particle can diffuse to its nearest 
neighbor at a rate satisfying the detailed balance condition. Now, we 
break up the lattice into cells of size $k \times k$, ( $k$ is any integer) and at low 
temperatures, a particle is only allowed to hop within its own cell. 
Then, in the low temperature phase, the number of particles in each cell 
gets frozen in time, and the number of such sectors grows exponentially with the volume of the system. However, 
calculating the pico-canonical partition functions in such models has 
not been possible so far.

If there is a sequence of ergodicity 
breaking transitions, (say corresponding to a smaller region $G_1$ inside $G$,
in Fig. 1), then the state will depend on  more parameters 
$\alpha_1^*, \alpha_2^*, \ldots$, and would show a much more complicated 
dependence on the history of the system.

DD would like to thank A. Ghosh and S. P. Das for some very useful 
discussions, R. Dickman for a critical reading of the paper, and the 
Government of India for financial support through a 
J. C. Bose Research Fellowship. The work of JLL was supported by 
NSF grant DMS-08-0212 and by AFOSR grant 095 50-07.

\end{document}